\documentstyle[epsf]{laa}     
 
\newcommand{\be}{\begin{equation}} 
\newcommand{\ee}{\end{equation}} 
\newcommand{\etal}{{\it et al.}} 
\newcommand{\hmp}{h^{-1}Mpc} 
\newcommand{\bef}{\begin{figure}} 
\newcommand{\eef}{\end{figure}}

\def\spose#1{\hbox to 0pt{#1\hss}} 
\def\ltapprox{\mathrel{\spose{\lower 3pt\hbox{$\mathchar"218$}} 
 \raise 2.0pt\hbox{$\mathchar"13C$}}} 
\def\gtapprox{\mathrel{\spose{\lower 3pt\hbox{$\mathchar"218$}} 
 \raise 2.0pt\hbox{$\mathchar"13E$}}} 
\def\inapprox{\mathrel{\spose{\lower 3pt\hbox{$\mathchar"218$}} 
 \raise 2.0pt\hbox{$\mathchar"232$}}} 
 
\begin{document} 
   \thesaurus{07         
              (07.09.1;  
               03.01.1;  
               04.01.1)  
             } 

\title{Comment on the paper   
  ``The ESO Slice Project galaxy redshift survey: 
V. Evidence for a D=3 sample dimensionality''} 
 
   \author{ M. Joyce \inst{1,2,3},
   M. Montuori \inst{1}, F. Sylos Labini \inst{1,4},   and  
   L. Pietronero \inst{1}} 
   \institute{        INFM Sezione Roma1,        
		      Dip. di Fisica, Universit\'a "La Sapienza", 
		      P.le A. Moro, 2,  
        	      I-00185 Roma, Italy. 
		\and 
		INFN Sezione Roma 1. 
                \and
                School of Mathematics, Trinity College, Dublin 2, Ireland.
                \and
                D\'ept.~de Physique Th\'eorique, Universit\'e de Gen\`eve,  
		24, Quai E. Ansermet, CH-1211 Gen\`eve, Switzerland.} 
 
\date{Received -- -- --; accepted -- -- --} 
 
\maketitle

\begin{abstract} 
In a recent analysis of number counts in the 
ESP survey  Scaramella et al. (1998) claim to find
evidence  for a cross-over to homogeneity at large
scales, and against a fractal behaviour with dimension
$D \approx 2$. In this comment we note firstly that, if 
such a cross-over exists as described by the authors,
the scale characterizing it is $\sim 100 \div 300 \hmp$.
This  invalidates the ``standard'' 
analysis of the same 
catalogue given elsewhere 
by the authors  which results in a ``correlation
length'' of only $r_0 = 4 \hmp$. 
Furthermore we show that the evidences for
a cross-over to homogeneity rely on the choice of
cosmological model, and most crucially on the
so called K corrections. 
We show that the $D \approx 3$
behaviour seen in the K-corrected data
of Scaramella et al. is in fact
unstable, increasing systematically towards $D=4$
as a function of the absolute 
magnitude limit. This behaviour
can be quantitatively explained as the effect of 
K-correction in the relevant range of red-shift ($z \sim 0.1
\div 0.3$). A more consistent interpretation of
the number counts is that $D$ is in the range $2 \div 2.5$,
depending on the cosmological model, consistent with 
the continuation of the fractal $D\approx 2$ behaviour
observed at scales up to $\sim 100 \hmp$.
This implies a smaller K-correction.
Given, however, the uncertainty in the effect of intrinsic
fluctuations on the number counts statistic, and
its sensitivity on these large scales to the uncertain 
K corrections, we conclude that it is premature 
to put
a definitive constraint on the galaxy distribution
using the ESP data alone.
\end{abstract}

\section{Introduction} 

In a recent paper Scaramella \etal (1998; S\&C ) have applied
the same statistical analysis to two deep 
surveys of large scale structure -the ESP galaxy survey, 
and the Abell and ACO survey of clusters - as performed by three of
us and reported in Sylos Labini \etal (1998; Paper 1).
Despite the adoption of the same method, they 
have reached very different conclusions: That there is no
evidence for a scale invariant distribution of galaxies
with dimensionality $D \approx 2$, as argued in Paper 1,
and that their results favour the ``canonical''
value of $D=3$ (corresponding to a homogeneous universe).
In this comment we will concentrate on a few specific 
points, referring the reader to Paper 1 for detailed 
discussion of various points on which we have nothing
new to add. Firstly we clarify what 
the claims of Paper 1 for a fractal distribution 
of galaxies actually are, in particular the strength of the 
statistical evidence on different scales, as the discussion given in 
S\&C is 
unclear on this point.  We then turn to a 
detailed discussion of the analysis of the ESP data,
discussing the results and different interpretations
of Paper 1 and S\&C. Our main substantial result is the
demonstration 
that the $D=3$ behaviour which appears
in the data when K corrected as prescribed in S\&C is
in fact unstable, with $D$ increasing  systematically 
to substantially larger values as a function of the 
depth of the volume limited sample. Quantitatively
this behaviour can be explained as the effect of  
a large K correction applied to
an underlying galaxy distribution with fractal 
dimension $D \approx 2$. 
We only briefly comment on the analysis of the Abell 
and ACO cluster catalogues as a full discussion 
of the relevant issues is given in Paper 1.

\section{Statistical Methods} 
The assumption of homogeneity in the distribution of matter
lies at the heart of the Big Bang cosmology. The
nature of the evidence, if any, for this assumption 
has, however, been the subject of 
considerable controversy
(Davis 1997, Pietronero \etal, 1997). 
A central point made by one of us (LP, Pietronero 1987) has been 
that the standard methods of analysis of galaxy red-shift 
catalogues, which provide the most direct probe of 
the (luminous) matter distribution,
actually assume homogeneity implicitly. 
An alternative analysis has been proposed using statistical methods 
which  do not make this assumption, and are appropriate
for characterizing the properties of regular as well as irregular
distributions. Essentially this analysis makes use of
a very simple statistic, the conditional
average density, defined as
\be 
\label{e4} 
\Gamma(r) = \left \langle \frac{1}{S(r)} \frac{dN(<r)}{dr} \right \rangle 
\ee 
where $dN(<r)$ is the number of points in a shell 
of radius $dr$ at distance $r$ from an occupied point and 
$S(r)dr$ is the volume of the shell. Note that in Eq.\ref{e4} 
there is an average over all the occupied points contained in the sample. 
If the distribution is scale invariant with fractal dimension $D$
we have
\be 
\label{ee5} 
\Gamma(r) \sim r^{-\gamma} \;  
\ee 
where $\gamma=3-D$. 

The problems of the standard
analysis can easily be seen from the fact that, for
the case of a fractal distribution
the standard ``correlation function'' $\xi(r)$ 
in a spherical sample of radius $R_s$, 
is given by 
\be 
\label{e7} 
\xi(r) = \frac{3-\gamma}{3}  
\left( \frac{r}{R_s} \right)^{-\gamma} -1 \; . 
\ee 
Hence for a fractal structure  the ``correlation length'' $r_0$ 
(defined by $\xi(r_0)=1$), is not a scale characterizing any
intrinsic property of the distribution, but just a scale
related to the size of the sample. If, on the other hand, the
distribution is fractal up to some scale $\lambda_0$ and
homogeneous beyond this scale, it is simple to show that,
 if $\lambda_0 < R_s$, i.e. the crossover to homogeneity 
is well inside the sample size, 
\be 
\label{e9} 
r_0 = \lambda_0 \cdot 2^{-\frac{1}{\gamma}} \; . 
\ee 
The correlation length does in this case have a real physical meaning
(when measured in samples larger than $\lambda_0$),
being related in a simple way to the scale characterizing
homogeneity. In the case $D=2$ we have $r_0 = \lambda_0/2$
\footnote{This calculation assumes a simple matching of
a fractal onto a pure homogeneous distribution. For 
any particular model with fluctuations away from 
perfect homogeneity, the numerical factor will differ
slightly depending on how precisely we define the 
scale $\lambda_0$.}.

\section{The debate on the fractal universe} 

The most popular point of view is that given, for example, by 
Peebles (1993) and Davis (1997) according to which
the homogeneity scale has already been well established by 
means of different observations. 
Peebles (1993) argues that 
the homogeneity of the matter distribution is indicated  by the 
uniformity of galaxy angular catalogs, the number counts of galaxy 
as a function of apparent magnitude and the 
so-called ``rescaling'' of the amplitude of the  
angular correlation function. In particular a  
correlation length $r_0 \approx 5 \hmp$ is 
taken to be a real physical scale characterizing the 
departure from  homogeneity. Increase in this quantity with 
depth of sample is ascribed to the so-called 
``luminosity bias'' (Davis \etal, 1988;  see 
also Benoist \etal, 1996). In complete disagreement 
with these claims is the direct analysis of all available
galaxy catalogues reported in Paper 1, using the 
conditional average density $\Gamma(r)$. Up to the deepest
scales for which such an analysis can be carried out
for available catalogues  clear evidence for fractal 
behaviour with $D \approx 2$ has been found. No evidence 
for  a cross-over to homogeneity at the scales required 
to make the standard analysis meaningful is found, nor 
indeed any evidence for such a cross-over at scales within
the range of current catalogues with robust statistics,
which place a lower
bound on $\lambda_0$ of approximately $100 \hmp$.
 
This length scale for these statements is fixed 
by the maximum radius of a sphere which can be fitted inside
the sample volume of the corresponding surveys, since the 
statistic $\Gamma(r)$ can only be computed up to such a distance.
To extract information from surveys 
about correlations
at length scales greater than this, one needs to consider
other statistics. The most simple one is the radial count 
$N(<r)$ from the origin in whatever solid angle is covered 
by the survey. Depending on the survey geometry the difference
between the length scales to which we can calculate $\Gamma(r)$
and $N(<r)$ varies greatly. The number count from the origin
is obviously a much less powerful statistic since it doesn't 
involve the average. It is intrinsically a more fluctuating 
quantity. Such fluctuations are, however, about the average 
behaviour and,  by sampling a sufficiently large range of scale,
one should be able to recover the average behaviour of the 
number count $\sim r^D$. What one means by ``sufficiently large range
of scale'' depends
strongly on the underlying nature of the fluctuations.
In particular (see section 6.1 of Paper 1) the
cases of a homogeneous distribution with Poissonian type
fluctuations and a fractal structure with scale invariant 
fluctuations are quite different. In the former case one 
predicts a very rapid approach to perfect $D=3$ behaviour
at a few times the scale characterizing the fluctuations;
in the latter $N(<r)$  has intrinsic fluctuations
on all scales (which average out in $\Gamma(r)$), 
in addition to the statistical 
sampling Poisson noise in $N(<r)$ which dominates
up to a scale which depends on the sample (on the number of points,
and therefore on the solid angle of the survey). 

\section{Analysis of the ESP redhsift survey}
 
We now turn to the ESP survey, which is a very deep survey 
extending to red-shifts $z \sim 0.3$ in a very narrow
solid angle $\Omega \approx 0.006 \; sr$. Because of this geometry 
the analysis with $\Gamma(r)$ only extends to $R_s \approx 10 \div 12 \hmp $;
in this regime it shows a clear $D\approx 2$ behaviour
consistent with the other surveys analyzed in Paper 1.
With the radial counts from the origin $N(<r)$, however,
the analysis can extend to distances almost two orders of 
magnitude greater. The results of the latter analysis
can be summarized as follows:

$\bullet$ Up to a scale of $\sim$ 300 $\hmp$ the number counts are
highly fluctuating;

$\bullet$ Beyond this scale the counts can be well fitted by a 
fairly stable average behaviour  $N(<r) \sim r^D$. 
\bigskip

The value 
obtained for the dimension of the number count $D$ 
in this range depends (i) on the assumptions about the
cosmology and (ii) on the so-called K corrections.
The uncorrected data (i.e. with the euclidean distance 
relation and without K corrections) gives a slope of 
$D \approx 2$, while both corrections lead to an increase in the slope.
In particular S\&C  apply the corrections in a way
which produces  values $D \approx 3$ which they argue to
reflect the true behaviour of the galaxy distribution\footnote{There is 
no significant disagreement in the results 
we obtain with a slightly earlier version of the catalogue
and those given in S\&C.  The flattening effect of the
corrections is somewhat obscured by figures 68 
and 70 in Paper 1 which included some extra points 
beyond the appropriate volume limit.}.

\subsection{The fluctuating regime: $r \ltapprox 300 \hmp$} 
  
Consider first what conclusions may be drawn from the
fluctuating regime. If the universe is homogeneous on 
large scales, the data clearly show that the scale 
characterizing such homogeneity is at least
$\sim 100 \div 300 \hmp$.
The implication from our discussion above is that
any standard analysis using the correlation function
is inappropriate for characterizing the properties of
the fluctuations.  In particular the authors claim 
in various papers (see e.g. Guzzo, 1997)  that
in the ESP galaxy sample, $r_0 \approx 4 \hmp$.
The origin of this length scale can also be
inferred from the discussion above:
In a fractal distribution (with $D=2$) we see from
(\ref{e7}) that $r_0 \approx R_s/3$, and as 
noted above $R_s \approx 10 \div 12$ in the ESP
survey. It is related not to a physical
property of the galaxy distribution but to the
specific geometry of the survey it has been
determined from. The more general implication of 
this observation of large fluctuations up
to 300 $\hmp$ for all standard analyses of existing 
catalogues is also clear
\footnote{For example, some of the authors of S\&C  have also worked 
on the SSRS2 galaxy sample, where they find $r_0 \approx 16 \hmp$ 
for the deepest volume limited (VL) sample, i.e. for  the most luminous 
galaxies (Benoist \etal, 1996). See Paper 1 for a discussion 
of the behaviour of $r_0$ with sample size in this survey
and Cappi et al. (1998) for a different interpretation.}.
If, on the other hand, the 
true underlying behaviour is fractal, the transition from
a highly fluctuating to a more stable behaviour can be understood
as the transition between large statistical fluctuations and
much smaller scale-invariant intrinsic fluctuations. In
Paper 1 it is shown how, using the dimension and 
normalization of the radial density from the analysis of 
the other surveys analyzed with the $\Gamma(r)$ statistic, 
one obtains a simple estimate for this scale in ESP which
agrees with the observation that it is $\sim$ 300 $\hmp$. 
The observed fluctuating regime is therefore consistent with 
the continuation of the fractal behaviour seen at smaller scales.

\subsection{The large scale regime: $r \gtapprox 300 \hmp$} 

Now let us move onto the conclusions which can be drawn 
from the smoother regime at larger scales. In Paper 1
it was noted that from $ \sim 300 \hmp$ the number counts 
show very approximately a dimension $D \approx 2$,
and therefore provide weak evidence for the continuation
of the behaviour seen at smaller scales. Different 
redshift-distance relations and some typical K corrections were 
also applied to the data, but, given the uncertainty 
in their determination, they were not interpreted as 
sufficiently reliable to change an already statistically 
weak conclusion. In particular, for example, detailed 
fits to different VL samples were not performed.
In S\&C the authors have looked at the precise effect which certain 
corrections can have on the data in the regime beyond $300 \hmp$
in much greater detail, and drawn the
strong conclusion that the $D \approx 2$ behaviour is ruled out
and $D=3$ behaviour clearly favoured. 

The difference of interpretation centres on the K corrections,
and the strength of any conclusion on the confidence with which
they can be made.  We will now see that some checking shows
that the K corrections as they have been applied to obtain this
result are 
actually 
inconsistent with the conclusion of
an underlying $D=3$ dimensionality. 
\bigskip

\subsection{Volume Limited (VL) samples and the effect of K-correction} 

In order to construct VL   catalogues we need to
determine the absolute magnitude $M$ of a galaxy at redshift 
$z$ from its observed apparent magnitude $m$, and it is here that
the K correction enters:
\be
\label{abs-mag} 
M = m -5 log (d_L) -25 -K(z)
\ee 
Here the luminosity distance $d_L=r(1+z)$
in Friedmann-Robertson-Walker (FRW) models where $r$ is the comoving distance,
and $d_L=(c/H_o)z$ in the euclidean case.
The K correction corrects for the fact that when a galaxy 
is red-shifted we observe it in a redder part of its spectrum 
where it may be brighter or fainter. It can in principle be determined
from observations of the spectral properties of galaxies,
and has been calculated for various galaxy types as
a function of red-shift (e.g. Fukugita \etal, 1995). 
When applying them to a red-shift
survey like ESP, we need to make various assumptions since
we lack information about various factors (galaxy types,
spectral information etc. as a function of red-shift
and magnitude).
 
It is not difficult to understand how K corrections
affect the number counts systematically. 
The number count in a 
volume limited sample corresponding to an
absolute  magnitude limit 
 $M_{lim}$ can be written schematically as
\be
\label{number-count}
N( < R)= \int_0^R \rho(r) d^3r \int_{- \infty}^{M_{lim}}  \phi(M,r) dM 
\ee
where $\rho(r)$ is the real galaxy density, and $\phi(M,r)$ the 
appropriately normalized luminosity function (LF) in the radial
shell at $r$. If the latter integral is independent
of the spatial coordinate (i.e. if the fraction of galaxies
brighter than a given absolute magnitude is independent
of $r$), it is just an overall normalization and
the exponent of the number counts in the VL   
sample show the behaviour of the true number count, 
with  $N(<R) \sim R^D$ corresponding
to the average behaviour $\rho \sim r^{D-3}$ as we have assumed.
What we do when we apply K corrections or other red-shift 
dependent alterations to the relation (\ref{abs-mag}) is 
effectively to 
change the LF $\phi(M,r)$. Using an incorrect relation 
will induce $r$ dependence in this function, which 
in turn will distort the relation between radial dependence of 
the number counts and the density. 

When one applies a K correction and observes a significant
change in the number counts, there are thus two possible
interpretations - that one has applied the physical
correction required to recover the underlying behaviour 
for the galaxy number density, 
{\it or} that one has distorted the LF to produce a 
radial dependence unrelated to the underlying density. 
How can one check which interpretation is correct? 
Consider the effect of applying a 
{\it too large}   K correction.
To a good approximation
the net effect is a linear shift in the magnitude $M$ with 
red-shift, so that $M \rightarrow M - kz$.
The second integral in (\ref{number-count}) can in this
case be written 
\be
\label{shifted-integral}
\int_{-\infty}^{M_{lim}} \phi_p (M+kz) dM = 
\int_{-\infty}^{M_{lim}+kz} \phi_p (M) dM 
\ee
where $\phi_p$ denotes the physical, r-independent LF. 
This is a function whose shape is well known to
be fitted by a very flat power-law with an exponential
cut-off at the bright end. If the upper cut-off of the
integral is in the former range, one can see from 
(\ref{number-count}), taking $z \sim r$, that 
the number count picks up an additional
contribution going as $R^{D+1}$, so that 
we expect the slope to increase by one 
at some sufficiently large scale.
As we go to the brighter end of the luminosity function,
where it turns over, the fractional number of galaxies 
being added by the correction is even greater and we 
expect to see a growing effect on the slope with depth 
of sample. To illustrate the effect quantitatively 
we have computed the integral (\ref{number-count}) 
numerically for a distribution $\rho(r)$ which has an
average $D=2$ behaviour in euclidean co-ordinates, 
taking $\phi (M,r)= \phi_p (M+K(z))$, where $\phi_p(M)$
is the LF for the ESP survey given in Zucca \etal (1997),
and $K(z)$ the average K correction used in S\&C.
\bef  
\epsfxsize 8cm  
\epsfysize 10cm   
\centerline{\epsfbox{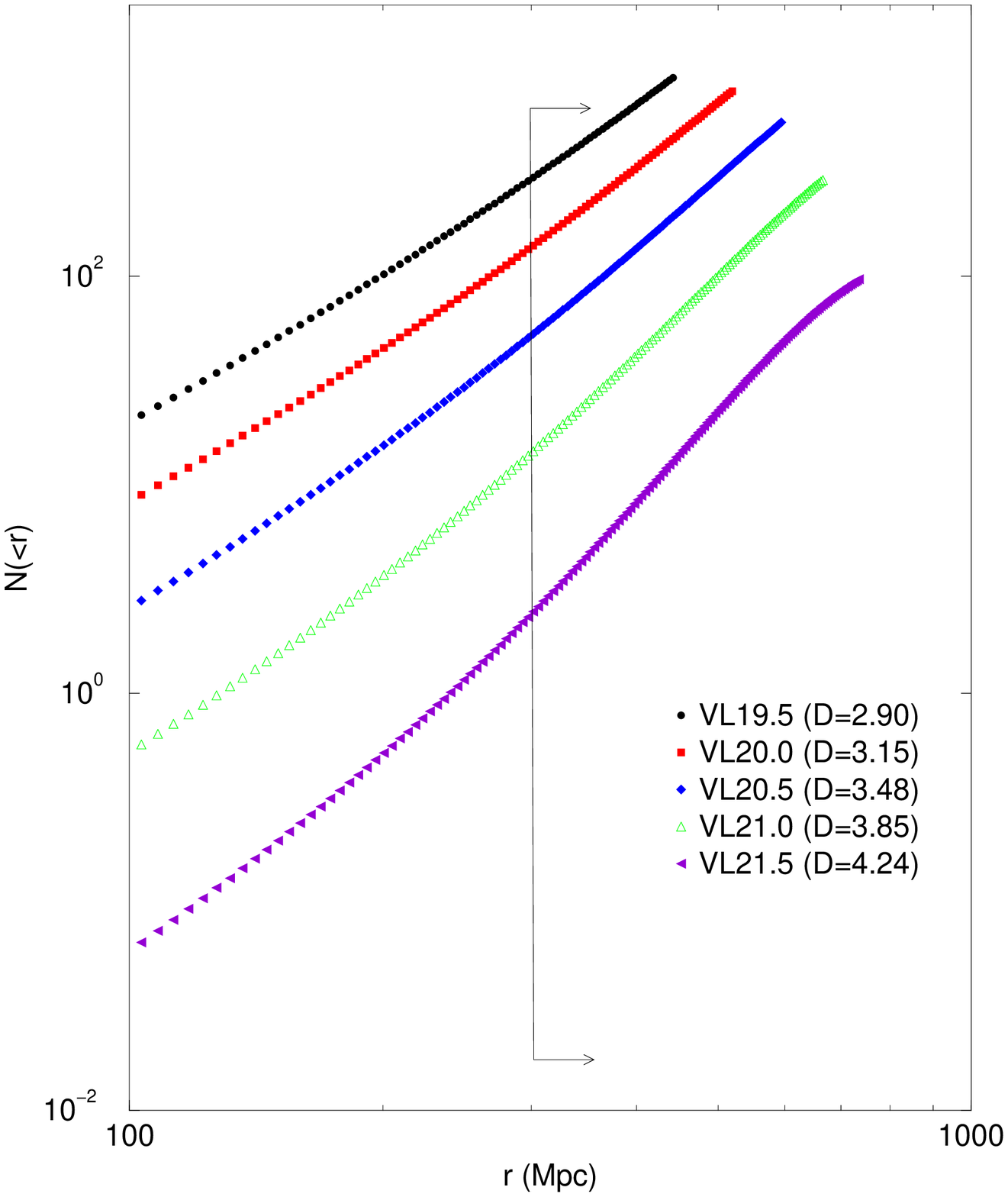}}  
\caption{A plot of average number counts 
in a survey with the LF of the ESP survey 
($\alpha=1.2$ and $M^*=-19.6$), if the 
underlying distribution has fractal dimension
$D=2$ in euclidean coordinates, with a 
shift to the LF corresponding to the K correction  
applied by S\&C.
The different VL samples are labelled
according to the cut in absolute magnitude
(i.e. VL20.5 corresponds to a cut at $M=-20.5$).
The distance $r$ is that for the flat FRW cosmology
used by S\&C ($q_o=0.5$). At scales  $r \gtapprox 300 \hmp$ 
we see that the effect is to produce an exponent
$D \approx 3$, increasing systematically to larger
values as we go to deeper VL samples with an
absolute magnitude cut $M \gtapprox M^*$. The fit with a power
law function has been done in the region to the right of
the arrow.
\label{kcorr}} 
\eef

The results plotted in Fig. \ref{kcorr} show that, on 
the scales over which the ESP data are analysed, 
we should be able to clearly distinguish the case of
a spurious K correction from the case of a real
underlying $D=3$ behaviour, which should be relatively
stable as a function of the absolute magnitude limit of
the VL sample. In Figure 2 of their paper S\&C show the 
number counts for a few VL samples,
and conclude that there is evidence for a real sample slope
of $D=3$, the variation being ascribed to random errors.
In Fig. \ref{scar2} we show precisely the same figure with additional
VL samples at greater depth, which have been omitted 
by S\&C\footnote{ Note that the 
sample with $M=-20.85$ in Fig. \ref{scar2}
has three times the number of galaxies in the 
$M=-20.5$ sample given in S\&C in their analysis of 
the uncorrected data. We have chosen to label our samples
primarily by red-shift, since this is the only relevant quantity for
a galaxy which does not change with corrections.}.
\bef 
\epsfxsize 8cm 
\epsfysize 10cm 
\centerline{\epsfbox{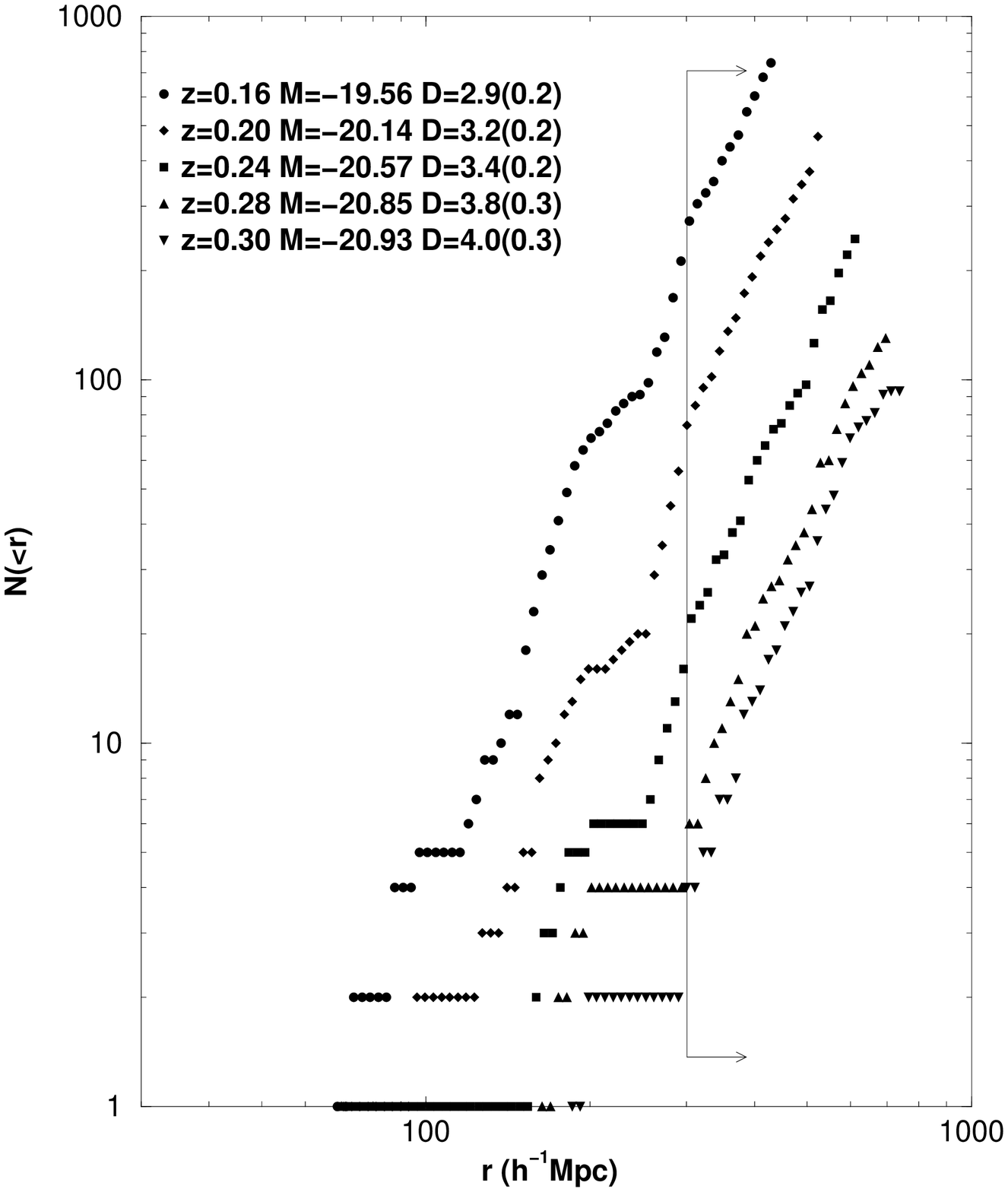}} 
\caption{ \label{scar2} 
The figure corresponds to Fig.2 of S\&C
($q_o=0.5$ with exactly the K correction given
in Zucca \etal  1997), but including deeper 
VL samples. 
For each VL sample it is reported the 
limiting redshift and absolute magnitude.
The systematic growth of the exponent 
of the number counts as a function of the 
absolute magnitude limit of the sample is clearly
seen. The fit with a power
law function has been done in the region to
 the rigth of
the arrow. The value of the fractal dimension $D$ 
for each VL sample 
is reported 
in the label; in brakets 
it is shown 
 the error on the fractal dimension
estimated with 40 boostrap resamplig.} 
\eef
The conclusions one can draw from these two figures 
are clear: The $D\approx 3$ behaviour observed by
S\&C in their K corrected data is clearly
not stable as would be the case if it represented the
real underlying behaviour of the density. On the 
contrary, as we see from Fig. \ref{kcorr}, the 
``corrected'' data are in fact clearly better interpreted
as indicating an underlying distribution with 
$D \approx 2$ which has been subjected to a  
too large K correction in the relevant range
of red-shift $z \sim 0.1 -0.3$.

\subsection{Dependence on the cosmological model} 

What about the dependence on cosmological model? Can we
find a more consistent $D=3$ fit with the same K corrections
by changing the deceleration parameter $q_o$? The effect of changing
cosmological model is two-fold: (i) It changes the 
relation between red-shift and the comoving distance in which
we should see the homogeneity, and (ii) it changes
the absolute magnitude through relation (\ref{abs-mag}).
The effect of the former is relatively minor for
the red-shifts we are considering here, increasing the
slope by at most about $0.2$. The latter effect 
has the same  form as that of the  K-correction 
since, at linear order in red-shift, 
$d_L = d_E[1+\frac{1}{2}(1-q_o)z+...]$ where
$d_E=(c/H_o)z$ is the euclidean distance,
and therefore from (\ref{abs-mag}) we see that
it is equivalent to an effective linear K correction with
$k_{eff}=5(1-q_o)/2 \ln 10$. So, taking any FRW model with
a sub-critical matter density essentially adds an even larger
K correction of the type taken by S\&C and leads
to steeper slopes with an even more unstable behaviour as
a function of depth. For example, we show the results in
Fig. \ref{kcorr-cosmconst} for a currently popular
critical FRW model with a cosmological constant $\Omega_\Lambda=0.7$
(giving $q_o=-0.55$)
\bef 
\epsfxsize 8cm 
\epsfysize 10cm 
\centerline{\epsfbox{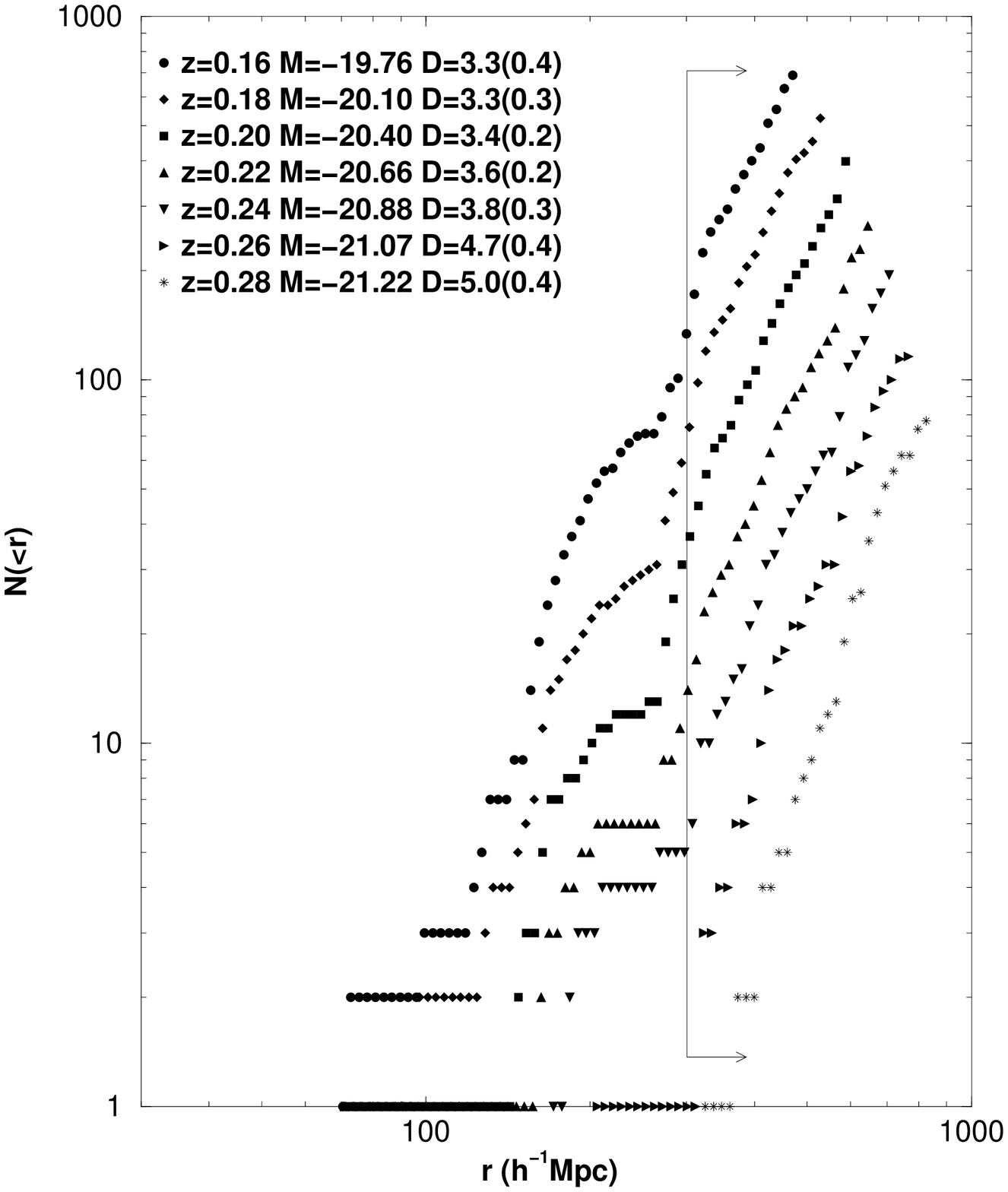}} 
\caption{ \label{kcorr-cosmconst} 
Same as Figure 2 but with $q_0=-0.55$
} 
\eef 
Adopting different FRW cosmological models from the one used
by S\&C in their analysis produces results even more 
inconsistent with 
homogeneity. 
In contrast with the conclusion of S\&C
that the $D \approx 2$ result of Paper 1 is only 
tenable by  ``both unphysically ignoring the galaxy
K-correction and using euclidean rather than FRW cosmological
distances'', we conclude therefore that the alternative $D=3$ result 
is arrived at only by applying an unphysical K correction 
and taking a  specific FRW cosmology.

The   strong  dependence on the chosen FRW cosmology 
is   interesting by itself. 
Without any K corrections the $q_o=0.5$ cosmology gives slopes
of $\approx 2.4$, while for the $q_o=-0.55$ case we get values
from $2.5$ to $2.8$ as we go to deeper samples. If  we had 
confidence in applying the K correction, very strong conclusions 
could be drawn about these models.  It is perhaps interesting
to note that any approximately linear K correction will
not produce a stable dimension near $D=3$ for this data in these 
models. Clearly the physical K correction in this range 
of red-shift must be a very non-trivial function quite 
different from that used by S\&C if the ESP data arises from
an underlying homogeneous distribution. 
In the absence of a clearly consistent and well understood way of
applying such corrections, it makes little sense to draw conclusions 
which depend so strongly on them. By contrast the interpretation
of a continuation of fractal behaviour with $D\approx 2$ (and a
relatively unimportant role for K corrections) is   consistent,
supported also by the behaviour seen in the 
fluctuating regime. With a smaller linear K correction with
k=1, for example, we find a stable slope in euclidean coordinates 
around $D=2.2$. 

\section{Discussion} 
 
Besides consistency requirements of this type, can we determine 
more directly what the physical K corrections are (in the
absence of direct measurements of the necessary spectral
information)? Clearly this is a general question which is not simply 
of relevance to ESP, but to the analysis of any survey which 
stretches to such scales. In particular, if there is a cross-over to
homogeneity, we have seen that it is at least at scales 
of hundreds of megaparsecs, and therefore these corrections will be 
of importance in the analysis of any forthcoming survey 
to search for  
evidence for homogeneous or fractal behaviour. The approach 
of S\&C takes the K corrections 
from other sources and applies them with various assumptions 
to their survey. We have seen that, at least for this case,
a simple check shows that this procedure 
gives incosistent results. 
Can we see this more directly? 
As we discussed above, the physical effect underlying 
the K correction 
produces
a red-shift dependent distortion of the (uncorrected) LF.
By looking at the (uncorrected) LF
in different slices of the survey we should, in principle,
be able to see  its effect.
 Furthermore, if we are applying K 
corrections to the data in the appropriate way, we should
be able to see them ``undo'' such distortion. 
For the ESP survey we have
broken up various VL limited samples into  
`nearby' and `distant' slices  and looked at the
 LF in the uncorrected and `corrected' catalogues.
This analysis shows that the uncorrected catalogue
shows good agreement between the LF in the two samples at 
different depth, compared to samples in the catalogue 
`corrected' as in S\&C, in which we see clear
evidence for a too large 
K-correction causing an unphysical distortion of the LF. 
However, because the statistics are weak,
we cannot 
derive any   strong   statement about what 
the real K correction effect is. 
An analysis  
 of the effect of K corrections on the LF should 
however provide very useful 
consistency checks in the analysis of the  much larger 
forthcoming redshift surveys.

\section{Conclusion} 
More generally we emphasize that the degree of
uncertainty in the results here is related to the
fact that we are using only the radial counts
from the origin, which are extremely sensitive to these
effects which produce systematic distortions relative
to this   origin. These effects will be much less  with
  the full correlation function, which averages over points.
In a forthcoming work we will discuss in more detail the effect
of different corrections on the various statistics.
To draw strong statistical conclusions about the 
distribution at large scales like those which have been
possible at moderate scales (up to $\sim$ 100$\hmp$)
 we will probably have to await the completion
  of the much larger 
surveys in progress. 

Finally a brief comment on the  
cluster distributions, of which a detailed discussion
is given in Paper 1.
An analysis with the conditional average density $\Gamma(r)$
is possible up to $\sim 80 \hmp$, and shows clearly
defined fractal properties with $D \approx 2$.
The number counts from the origin show a quite fluctuating
behaviour up to about $\sim 100 \hmp$, and clear
incompleteness at scales considerably larger than this
(evidenced by a steep decrease in the number counts). 
The identification of a range of scale in which
the catalogue is `complete'  is itself strongly 
dependent on the assumption of homogeneity (as can
be seen, for example, from Figure 4 showing the raw
cluster data and the discussion of it 
in Scaramella \etal, 1991 ). In our
view the very good fit to a $D=3$ behaviour 
presented again in S\&C (based on the analysis
in Scaramella \etal, 1991) is an artefact of 
this assumption rather than the identification
of any real property of the spatial distribution 
of clusters.


\section*{Acknowledgments} 
We thank again Dr. Paolo Vettolani for having given 
us the possibility to analyze a preliminary version 
of the ESP data, and 
to  publish the results  
before the publication of the catalog. 
F.S.L. warmly thanks Y.V. Baryshev,  
P, Teerikorpi, and A. Yates  
for very useful discussions,  
remarks and comments. 
We also thank  the anonymous referee
and Dr. J. Lequeux for 
many suggestions and comments which  have improved
the clarity of the paper.
This work has been partially supported by the  
EEC TMR Network  ``Fractal structures and   
self-organization.''  
\mbox{ERBFMRXCT980183}  and by the Swiss NSF.


\begin{thebibliography}{99} 
  
\bibitem
{ben96} Benoist C., \etal  1996  
Astrophys. J. 472, 452 
 
\bibitem
{cp92}   
Coleman, P.H. \& Pietronero, L., 1992 Phys.Rep. 231,  311   

\bibitem
{cappi98}   
Cappi A. et al., 1998 Astron.Astrophys. 335, 779   

 
\bibitem
{dav97}Davis, M.,  in the Proc of the  
Conference "Critical Dialogues in Cosmology" N. Turok Ed. 1997 
World Scientific 
 
\bibitem
{dav88} Davis M. \etal, 1988 Astrophys. J. Lett.  333 L9 
 
\bibitem
{gu97} Guzzo L., 1997 New Astronomy 2, 517 
 
\bibitem{fuk95} 
Fukugita M., Shimasaku K. and  
Ichikawa T.   1995 PASP 107, 945   
 
 
 
\bibitem
{pee93}Peebles P. J. E., 1993 
Principles of physical cosmology, Princeton Univ. Press 
 
 
\bibitem
{pie87} Pietronero L. 1987 Physica A, 144, 257  
 
\bibitem
{pmsl97} Pietronero L., Montuori M. and   Sylos Labini F.
in the Proc of the  
Conference "Critical Dialogues in Cosmology" N. Turok Ed. 1997 
World Scientific 
 
\bibitem
{sca98}Scaramella R. \etal, 1998  
Astron.Astrophys. 334, 404 (S\&C). 
 
\bibitem
{svz}Scaramella R. \etal,  
1991, Astron. J. 101, 342.  
  
  
\bibitem
{slmp98} Sylos Labini F., Montuori M.,  
 Pietronero L., 1998  Phys.Rep. 293, 66 (Paper 1) 
 
 
 \bibitem
{vet97} Vettolani \etal,1997  Astron.Astrophys. 325, 954 
  
\bibitem%
{zucca97} Zucca E., \etal, 1997 Astron.Astrophys. 326,477  
 \end{thebibliography}
\end{document}